\documentclass[fleqn,usenatbib,useAMS]{mnras}
\pdfoutput=1
\usepackage{mathptmx}

\usepackage[T1]{fontenc}
\usepackage{ae,aecompl}

\usepackage{graphicx}	
\usepackage{amsmath}	
\usepackage{amssymb}	

\newcommand{\kms}{\ensuremath{\mathrm{km}\,\mathrm{s}^{-1}}}

\newcommand{\hi} {{\rm H\,{\footnotesize\rm I}}}

\newcommand{\ci} {{\rm [C\,{\footnotesize\rm I}]}}
\newcommand{\cii} {{\rm [C\,{\footnotesize\rm II}]}}
\newcommand{\nii} {{\rm [N\,{\footnotesize\rm II}]}}
\newcommand{\oiii} {{\rm [O\,{\footnotesize\rm III}]}}

\title[Neutral versus ionized gas kinematics at $z\simeq2.6$]{Neutral versus ionized gas kinematics at $z\simeq2.6$:\\The AGN-host starburst galaxy PKS\,0529-549}

\author[F. Lelli et al.]{
Federico Lelli,$^1$\thanks{ESO Fellow. E-mail: flelli@eso.org}, Carlos De Breuck$^1$, Theresa Falkendal$^{1,2}$, Filippo Fraternali$^3$,
\newauthor Allison W. S. Man$^1$, Nicole P. H. Nesvadba$^4$, Matthew D. Lehnert$^2$ \\
$^{1}$European Southern Observatory, Karl-Schwarschild-Strasse 2, 85748 Garching bei M\"{u}nchen, Germany\\
$^{2}$Sorbonne Universit\'{e}, CNRS UMR 7095, Institut d'Astrophysique de Paris, 98 bis bd Arago, 75014 Paris, France\\
$^{3}$Kapteyn Astronomical Institute, University of Groningen, Postbus 800, 9700 AV, Groningen, The Netherlands \\
$^{4}$Institut d'Astrophysique Spatiale, UMR 8617, Universit\'{e} Paris-Sud, B\^{a}t. 121, 91405, Orsay, France
}

\date{Accepted 2018 June 28; Received 2018 June 04; in original form 2018 March 05}

\pubyear{2018}

\begin{document}
\label{firstpage}
\pagerange{\pageref{firstpage}--\pageref{lastpage}}
\maketitle

\begin{abstract}
We present a multiwavelength study of the AGN-host starburst galaxy PKS\,0529-549 at $z\simeq2.6$. We use (1) new ALMA observations of the dust continuum and of the \ci\ 370 $\mu$m line, tracing molecular gas, (2) SINFONI spectroscopy of the \oiii\,5007\,\AA\ line, tracing ionized gas, and (3) ATCA radio continuum images, tracing synchrotron emission. Both \ci\ and \oiii\ show regular velocity gradients, but their systemic velocities and position angles differ by $\sim$300 \kms\ and $\sim$30$^{\circ}$, respectively. The \ci\ is consistent with a rotating disc, aligned with the dust and stellar continuum, while the \oiii\ likely traces an outflow, aligned with two AGN-driven radio lobes. We model the \ci\ cube using 3D disc models, which give best-fit rotation velocities $V_{\rm rot}\simeq310$ \kms and velocity dispersions $\sigma_{\rm V}\lesssim30$ \kms. Hence, the \ci\ disc has $V_{\rm rot}/\sigma_{\rm V} \gtrsim 10$ and is not particularly turbulent, similar to local galaxy discs. The dynamical mass ($\sim$10$^{11}$ M$_{\odot}$) is comparable to the baryonic mass within the errors. This suggests that baryons dominate the inner galaxy dynamics, similar to massive galaxies at $z\simeq0$. Remarkably, PKS\,0529-549 lies on the local baryonic Tully-Fisher relation, indicating that at least some massive galaxies are already in place and kinematically relaxed at $z\simeq2.6$. This work highlights the potential of the \ci\ line to trace galaxy dynamics at high $z$, as well as the importance of multiwavelength data to interpret gas kinematics.
\end{abstract}




\section{Introduction}

Over the past decade much progress has been made in the study of gas dynamics at high redshifts. Integral field spectroscopy (IFS) in the near-infrared (NIR) has been widely used to trace the ionized gas kinematics in star-forming galaxies at $z \simeq 1-3$ \citep[e.g.,][]{Forster2009, Gnerucci2011, Wisnioski2015, Stott2016, Nesvadba2017}. Radio and sub-mm observations with the Jansky Very Large Array and the Plateau de Bure Interferometer have proved to be powerful in tracing CO kinematics up to $z\simeq4$ \citep[e.g.,][]{Ivison2012, Hodge2012, Tacconi2013}. Moreover, the Atacama Large Millimeter Array (ALMA) has opened a new window to study galaxy dynamics at even higher redshifts ($z\simeq 4-7$) using the \cii\ emission line \citep[e.g.,][]{DeBreuck2014, Jones2017, Shao2017, Smit2017}.

These different studies probe different gas phases. NIR spectroscopy probes emission lines like H$\alpha$ and \oiii, tracing warm ionized gas ($\sim10^{4}$ K). Radio and sub-mm observations, instead, probe CO, \ci, and \cii\ transitions, tracing cold neutral gas ($<100$ K) like molecular hydrogen (H$_2$).

In nearby galaxies, H$_2$ and \hi\ generally display regular disc rotation, whereas H$\alpha$ and \oiii\ kinematics may be more complex. For example, in galaxies hosting starbursts and/or active galactic nuclei (AGN), H$\alpha$ and \oiii\ are often associated with galactic winds \citep[e.g.,][]{Arribas2014, Harrison2014}. In high-$z$ galaxies, it is unclear how these different gas phases relate to each other and whether they share the same kinematics, since only a few galaxies have been studied using multiple gas tracers \citep{Chen2017, Ubler2018}. In this context, the redshift window $z\simeq2-3$ is very interesting because (1) both ionized and neutral gas can be observed and spatially resolved with existing facilities, and (2) it corresponds to the peak of the cosmic star formation history \citep{Madau2014}, a key epoch to understand the formation of massive galaxies.

Here we compare neutral and ionized gas kinematics in the starburst galaxy PKS\,0529-549 at $z\simeq2.6$, combining new \ci\,$(2-1)$ data from ALMA with IFS data from SINFONI \citep{Nesvadba2017}. PKS\,0529-549 is a well-studied radio galaxy with an exquisite set of ancillary data: optical spectroscopy, ground-based NIR imaging, and radio polarimetry \citep{Broderick2007}, $Spitzer$/IRAC imaging \citep{deBreuck2010}, and 1.1 mm data from AzTEC \citep{Humphrey2011}. PKS\,0529-549 hosts a Type-II AGN and shows two radio lobes. The Eastern lobe holds the record for the highest Faraday rotation measure to date, implying a strong magnetic field and/or a dense circum-galactic medium \citep{Broderick2007}. With only 5 minutes of ALMA on-source integration time, we spatially resolve the \ci\ emission of PKS\,0529-549. We show that \ci\ and \oiii\ display different kinematics and discuss possible interpretations with the aid of 3D kinematic models.

Throughout this paper, we assume a flat $\Lambda$ cold dark matter ($\Lambda$CDM) cosmology with $H_{0}= 67.8$ km s$^{-1}$ Mpc$^{-1}$, $\Omega_{\rm m} = 0.308$ and $\Omega_{\Lambda} = 0.692$ \citep{Planck2016}. In this cosmology, 1 arcsec = 8.2 kpc at $z= 2.6$.

\section{Data Analysis}

\subsection{ALMA data}\label{sec:ALMA}

PKS\,0529-549 was observed on 2 September 2014 during ALMA Cycle 2 with 34 working antennas, giving maximum and minimum baselines of 1090 m and 33 m, respectively. The on-source integration time was 5 minutes. The flux and bandpass calibrator was J0519-4546, while the phase calibrator was J0550-5732. We used four 1.875\,GHz-wide spectral windows centered at 224.99, 226.87, 239.99 and 241.87\,GHz. The second spectral window cointains the redshifted $\ci\,(2-1)$ line, which corresponds to the $^3P_2 \rightarrow ^3P_1$ transition and has a rest-frame frequency of 809.3435 GHz.

The data reduction was performed using the Common Astronomy Software Applications (CASA) package \citep{McMullin2007}. The $uv$ data were flagged and calibrated using the standard CASA pipeline. The imaging was performed using the task $clean$ with $Robust=2$, which corresponds to natural weighting of the visibilities. This provides the highest sensitivity but the lowest spatial resolution. The resulting synthesized beam is $0.43'' \times 0.28''$ with a position angle (PA) of 64.5$^{\circ}$. A continuum image was obtained combining all four spectral windows, reaching a sensitivity of 0.055 mJy beam$^{-1}$. The \ci\ cube was imaged using a channel width of 50 \kms, giving a sensitivity of 0.55 mJy beam$^{-1}$ per channel. The continuum was subtracted from the \ci\ cube in the image plane, after fitting a first-order polynomial to line-free channels. We also attempted continuum subtraction in the $uv$ plane (before the Fourier transform), but sanity checks revealed that the continuum was over-subtracted in this way. In any case, the continuum subtraction does not strongly affect our kinematic analysis. The \ci\ flux is $2.1\pm0.4$ Jy\,\kms, as estimated from fitting the global \ci\ profile (presented in A. Man et al. 2018, in prep.).

After imaging, the \ci\ cube was analysed using the Groningen Imaging Processing System \citep[Gipsy,][]{Vogelaar2001}. To build the total \ci\ map (moment zero), we sum the signal within a Boolean mask, which follows the kinematic structure of the \ci\ emission. Given the low signal-to-noise (S/N) ratio, the mask is constructed interactively on a channel-by-channel basis by defining regions in which contiguous pixels have S/N\,$\gtrsim$\,2. When using a mask, the noise in the total map is not uniform because a different number of channels is summed at each spatial pixel. It is however possible to construct S/N maps and define a pseudo-3$\sigma$ contour by considering spatial pixels with S/N ratio between 2.75 and 3.25 \citep[see Appendix in][]{Lelli2014c}. We measure pseudo-$3\sigma \simeq 150$ mJy\,beam$^{-1}$\,km\,s$^{-1}$.

To build the velocity map (moment one), we estimate an intensity-weighted mean velocity for the pixels inside the Boolean mask. We stress that the velocity map is very uncertain due to beam smearing effects and the low S/N ratio. This map merely provides a rough overview of the gas kinematics. We do not consider the dispersion map (moment two) because it is even more strongly affected by beam smearing: moment-two values are not indicative of the intrinsic gas velocity dispersion because the \ci\ line is largely broadened by the low spatial and spectral resolutions. Our kinematic analysis uses the full 3D information and takes resolution effects into account: we build model cubes and compare them with the observed one using Position-Velocity (PV) diagrams and channel maps (Sect.\,\ref{sec:disc}).

\subsection{SINFONI and ATCA data}

\citet{Nesvadba2017} present IFS data of PKS\,0529-549 using SINFONI at the ESO Very Large Telescope (VLT). Two different datacubes were obtained: one at low spatial resolution ($1.2''\times1.2''$) where the stellar continuum and several emission-lines are detected, and one at high spatial resolution ($0.7''\times0.6''$) where only the strongest emission lines (\oiii$\lambda$4958, \oiii$\lambda$5007, and H$\alpha$) are recovered. The H$\alpha$ line is blended with the \nii\ doublet, creating a broad ($\gtrsim$100 \AA) and asymmetric line. We refer to \citet{Nesvadba2017} for further details.

We use the Gipsy package to analyse the high-resolution cube. To increase the S/N ratio of the emission lines, the cube was box-averaged over 3$\times$3 pixels. We constructed emission-line maps by summing the signal within narrow wavelength ranges, which were determined by visual inspection excluding channels contaminated by sky lines. No continuum subtraction was attempted, but we used the low-resolution datacube to estimate that the net continuum contribution on the emission-line maps is negligible. 
\label{fig:Atlas}
We constructed velocity fields by fitting a Gaussian function at each spatial pixel. This is preferable to intensity-weighted mean velocities given the relatively high S/N ratio. Each emission line was fitted independently considering a narrow wavelength range and rejecting fit results with unphysically small amplitudes ($\lesssim$5 counts) and/or dispersions ($\lesssim$5 \AA). Given the overall data quality, it would be untrustworthy to fit multiple Gaussians to try de-blending the H$\alpha$ and \nii\ lines. Hereafter, we consider only the \oiii$\lambda$5007 velocity field because the H$\alpha$+\nii\ and \oiii$\lambda$4958 ones have lower quality due to the lower S/N ratio. These emission lines, however, display similar kinematics.

The absolute positional accuracy of the SINFONI data relative to the World Coordinate System is $\sim$1$''$ \citep[see][for details]{Nesvadba2017}. Since the ALMA positional accuracy is much higher than 1$''$, we improve the SINFONI astrometry using the following approach. It is sensible to assume that the SINFONI stellar continuum and the ALMA dust continuum are physically related: the two maps indeed display a similar elongated morphology with a PA$\simeq$75$^{\circ}$ (Fig.\,\ref{fig:Atlas}). Thus, we overlay the emission peak of the SINFONI stellar continuum with that of the ALMA dust continuum. The resulting spatial shift is smaller than 1$''$.

We also consider radio continuum images from the Australia Telescope Compact Array (ATCA), which were provided by \citet{Broderick2007}. These authors published natural-weighted images at various frequencies and resolutions. Here we show the uniform-weighted image at 18.5 GHz because it has the highest spatial resolution ($0.55''\times0.37''$). This image is published here for the first time. 

\begin{figure*}
\centering
\includegraphics[width=0.93\textwidth]{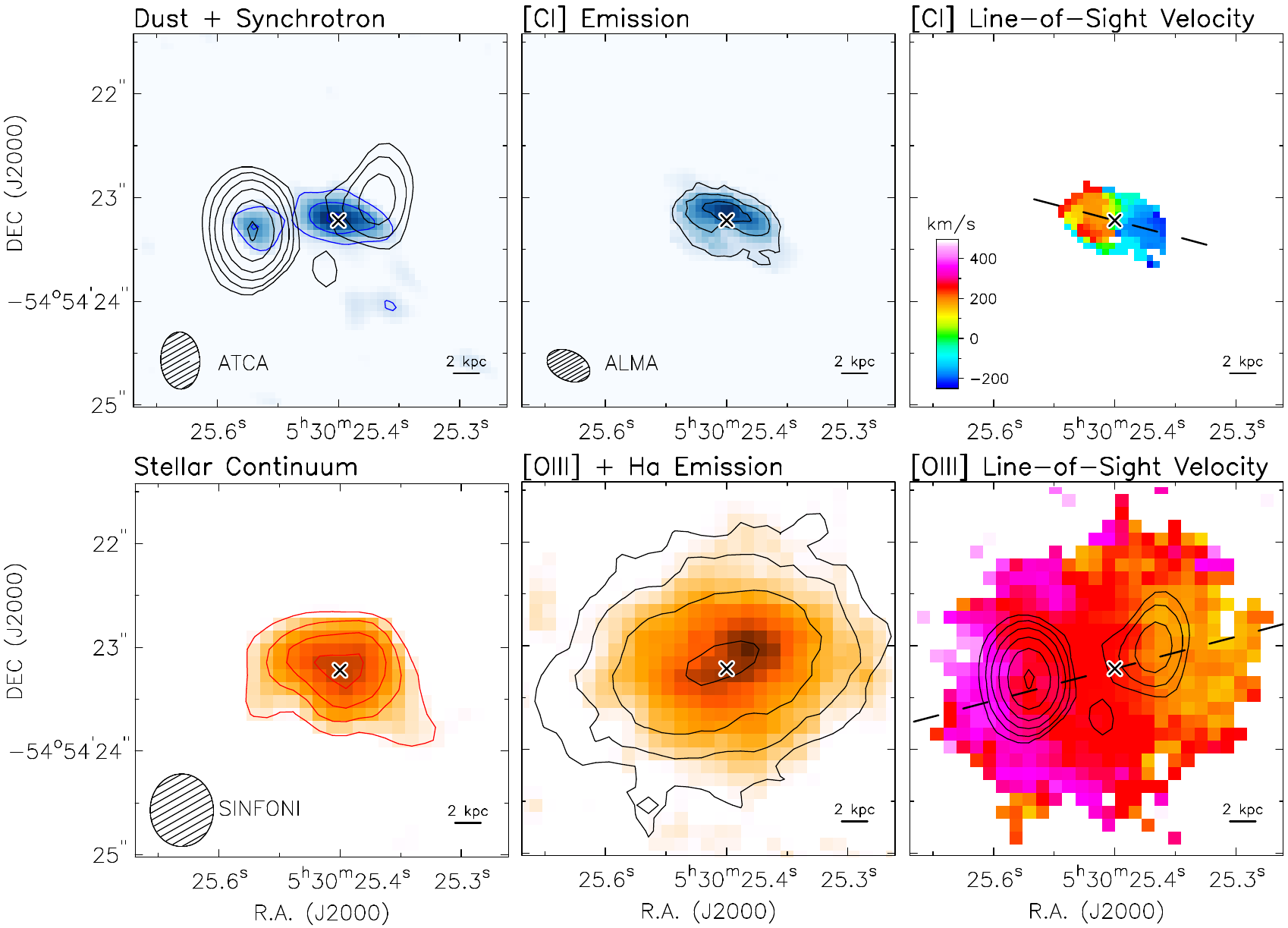}
\caption{Multiwavelength overview of PKS\,0529-549. \textit{Top left}: ALMA continuum (blue contours and colorscale) overlaid with the ATCA continuum at 18.5 GHz (black contours). The central ALMA component traces dust, while the Eastern component traces synchroton emission coinciding with the radio lobe. Blue contours are at 0.16, 0.32, 0.64 mJy beam$^{-1}$. Black contours are at 0.5, 1, 2, 4, 8, 16 mJy beam$^{-1}$. The ellipse shows the ATCA beam. \textit{Top middle}: Total [CI] map. Countours are at 150, 300, and 450 mJy\,beam$^{-1}$\,km\,s$^{-1}$. The ellipse shows the ALMA beam. \textit{Top right}: [CI] velocity field. The dashed line shows the [CI] kinematical PA. \textit{Bottom left}: Stellar continuum from SINFONI. The ellipse shows the PSF of the high-resolution SINFONI cube. \textit{Bottom middle}: H$\alpha$+[NII] emission (red colorscale) overlaid with the [OIII]$\lambda$5007 emission (black contours). \textit{Bottom right}: [OIII]$\lambda$5007 velocity field overlaid with the ATCA image at 18.5 GHz (black contours). The velocity scale is the same as in the top-right panel. The dashed line shows the [OIII]$\lambda$5007 kinematical PA. In all panels, the cross shows the [CI] centre, while the bar to the bottom right corner corresponds to 2 kpc.}
\label{fig:Atlas}
\end{figure*}

\begin{figure*}
\centering
\includegraphics[width=0.93\textwidth]{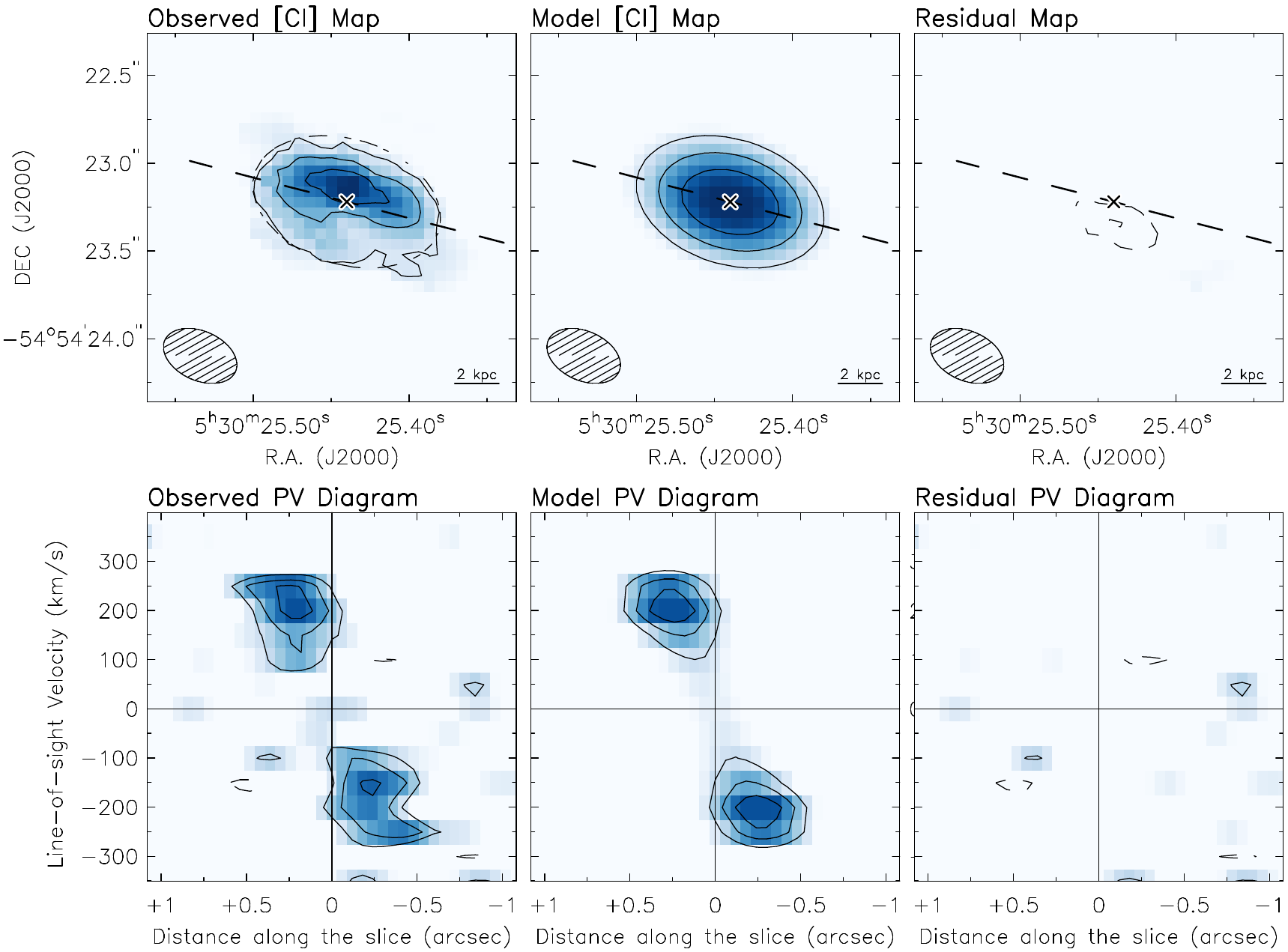}
\caption{\textit{Top panels}: observed [CI] map (\textit{left}), model [CI] map (\textit{middle}), and residual map (\textit{right}). Countours are at 150, 300, and 450 mJy\,beam$^{-1}$\,km\,s$^{-1}$. In the left panel, the dashed line shows the 3$\sigma$ countour from the model map. In all panels, the cross, the dashed line, and the ellipse show the disc center, the major axis, and the ALMA beam, respectively. \textit{Bottom panels}: Major-axis PV diagrams obtained from the observed (\textit{left}), model (\textit{middle}), and residual (\textit{right}) cubes. Contours are at $-$2$\sigma$ (dashed), 2$\sigma$ (solid), 3$\sigma$, and 4$\sigma$ where $\sigma=0.55$ mJy beam$^{-1}$. The vertical and horizontal lines correspond to the dynamical center and systemic velocity, respectively.}
\label{fig:Models}
\end{figure*}

\section{Results}\label{sec:results}

\subsection{Multiwavelength overview}

Figure\,\ref{fig:Atlas} provides a multiwavelength view of PKS\,0529-549. The top panels present the ALMA data. The ALMA continuum (left panel) is studied in detail in T.\,Falkendal et al. (2018, in prep.) by modelling its spectral energy distribution (SED). It consists of two components: a central one due to dust emission and a secondary one to the East due to synchrotron emission. The secondary component, indeed, nicely overlaps with the Eastern radio lobe from ATCA observations. The \ci\ emission (middle panel) coincides with the dust emission and displays a similar elongated morphology with $\rm{PA}\simeq75^{\circ}$. Both dust and \ci\ emissions are spatially resolved with $\sim$2 beams along the major axis, so their morphology is not strongly driven by the elongated beam shape, which has a similar PA of $\sim64.5^{\circ}$. No \ci\ emission is associated with the synchrotron lobes. The \ci\ velocity field (right panel) shows a velocity gradient along the major axis: this is consistent with a rotating disc as we show in Sect.\,\ref{sec:disc}.

The bottom panels of Fig.\,\ref{fig:Atlas} present the SINFONI data. The stellar continuum (left panel) is barely resolved but it seems to display roughly the same PA as the dust and \ci\ emissions. The ionized gas (middle panel) appears more extended than any other galaxy component. The SINFONI point-spread function (PSF) is significantly larger than the ALMA beam, but if we smooth the ALMA maps to the same spatial resolution ($0.7''\times0.6''$), they remain less extended than the ionized gas. The \oiii$\lambda$5007 velocity field (right panel) displays a velocity gradient with PA$\sim105^{\circ}$, so it is tilted by $\sim30^{\circ}$ with respect to the \ci\ distribution and kinematics. The \oiii\ velocity gradient, instead, aligns well with the orientation of the two synchrotron lobes. Strikingly, the \oiii\ velocities are systematically higher than the \ci\ velocities by $\sim$300 \kms.

The H$\alpha$ displays similar kinematics as the \oiii\, but its velocity field (not shown) is very uncertain due to the low S/N ratio and contamination from the \nii\ lines. In A.\,Man et al. (2018, in prep.), we present an X-Shooter spectrum of PKS\,0529-549 revealing that the \oiii\ and H$\alpha$ velocities are consistent with other emission lines from ionized gas like [OII] and CII], while the \ci\ velocities are consistent with photospheric absorption lines from young stars.

Given the evidence above, we interpret the \ci\ emission as a rotating gas disc, having a similar PA as both the dust component and the stellar continuum. The \oiii\ and the H$\alpha$ emission, instead, could trace a gas outflow, being aligned with the synchrotron lobes. \citet{Nesvadba2017} provide additional arguments for the outflow interpretation. For example, the \oiii$\lambda$5007 profiles are very broad, ranging from $\sim$300 to $\sim$900 \kms\ (see their Figure A.5): this is likely due to the contribution of different velocity components along the line of sight as expected in a gas outflow. The velocity difference between \ci\ and ionized gas may indicate that the outflow does not propagate exactly from the center of the \ci\ disc, but from its North-Eastern, redshifted side. Alternatively, we may be seeing only the redshifted side of the flow (further away from the observer) because the ionized bi-cone may have asymmetric density distribution, as observed in some local starburst galaxies \citep[e.g., NGC\,253,][]{Boomsma2005} and high-$z$ ones \citep[e.g., TN\,J1338-1942,][]{Zirm2005}. Interestingly, if only SINFONI data were available, it would be much more difficult to distinguish the \oiii\ velocity field of an outflow from that of a rotating disc. This highlights the importance of having multiwavelength observations to interpret ionized gas data. This point is further discussed in Sect.\,\ref{sec:compa}.

\begin{figure*}
\includegraphics[width=0.93\textwidth]{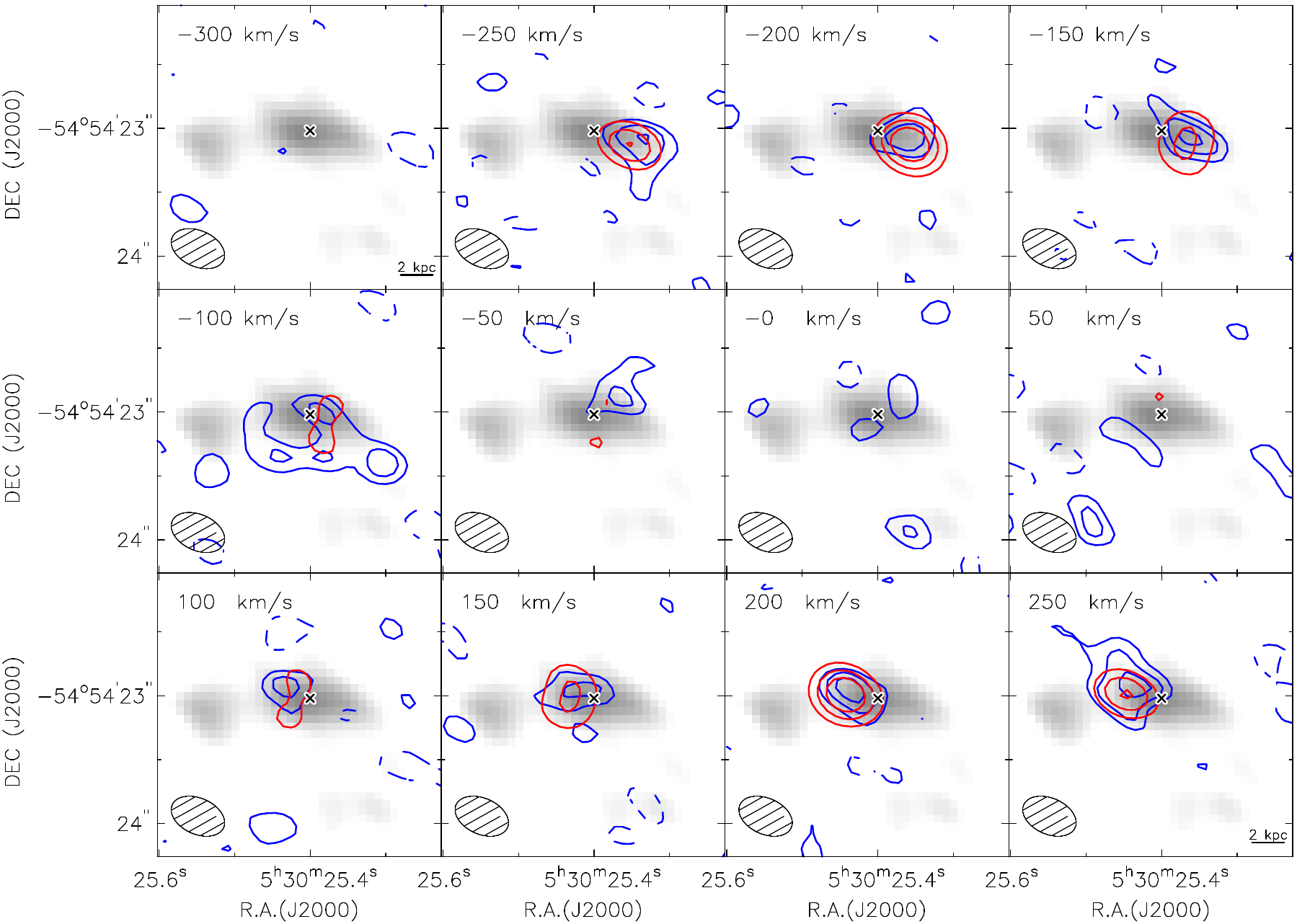}
\centering
\caption{[CI] channel maps from the observed (blue) and model (red) datacubes overlaid on the ALMA continuum (greyscale). Contours are at $-$2$\sigma$ (dashed), 2$\sigma$ (solid), 3$\sigma$, and 4$\sigma$ where $\sigma=0.55$ mJy beam$^{-1}$. In the top-left corner, we provide the line-of-sight velocity with respect to $V_{\rm sys}$. The ALMA beam is shown in the bottom-left corner. The cross corresponds to the dynamical center.}
\label{fig:ChanMaps}
\end{figure*}

\subsection{Disc modelling}\label{sec:disc}

To test the hypothesis of a rotating \ci\ disc, we use two different codes: GALMOD in Gipsy \citep{Sicking1997} and $^{\rm 3D}$BAROLO \citep{DiTeodoro2015}. Both codes build 3D kinematic models of rotating discs, taking projection and resolution effects into account. The main difference between the two is that GALMOD needs to be used iteratively by trial and error, while $^{\rm 3D}$BAROLO explores the full parameter space to find a best-fit solution by $\chi^{2}$ minimization. In practice, we use GALMOD to set the initial estimates of $^{\rm 3D}$BAROLO and to fix several geometric parameters that are otherwise degenerated.

The disc is built using a set of circular rings. Each ring is characterized by four projection parameters (centre, PA, inclination $i$, and systemic velocity $V_{\rm sys}$) and four physical ones (surface density $\Sigma$, vertical scale height $h$, rotation velocity $V_{\rm rot}$, and velocity dispersion $\sigma_{\rm V}$). The disc is then projected onto the sky to generate model cubes that are smoothed to the same spatial and spectral resolutions of the observations. Since the \ci\ distribution is resolved with only $\sim$2 beams, we assume that the disc parameters are constant with radius ($R$). This is equivalent to using a single, large ring. Effectively, the physical parameters are intensity-weighted average values over the semi-major axis of the \ci\ disc.

We explored two basic models: a disc with uniform surface density (the ring extends from $R=0$ to $R\simeq0.45''$) and a disc with a central hole (the ring extends from $R\simeq0.20''$ to $R\simeq0.45''$). Both models give comparable results because the ALMA beam ($0.43'' \times 0.28''$) smooth out any intrinsic structure in the gas density. This degeneracy only affects the determination of $\Sigma$, which cannot be robustly estimated and is not used in our analysis. For similar reasons, the model is insensitive to the vertical density distribution: we assume an exponential vertical profile with fixed $h = 0.04'' \simeq 300$ pc.

We use GALMOD following \citet{Lelli2015}: we inspect channel maps and PV diagrams to obtain initial estimates of the disc parameters, which are then refined until we find a good match between model and observations. The center, PA, and $V_{\rm sys}$ are easy to estimate and kept fixed during subsequent iterations. The remaining parameters ($i$, $V_{\rm rot}$, and $\sigma_{\rm V}$) are more challenging to determine due to degeneracies. The inclination is estimated by comparing model \ci\ maps (after beam convolution) to the observed one. Figure\,\ref{fig:Models} (top panels) shows that the observed \ci\ map is well reproduced by our model map: the residuals are smaller than $\sim$3$\sigma$ and less extended than the beam. These residuals are probably due to deviations from a pure axisymmetric gas distribution in the central regions. Clearly, the \ci\ disc is not face-on ($i>30^{\circ}$), so the net correction on $V_{\rm rot}$ is relatively small going as $1/\sin(i)$.

Finally, $V_{\rm rot}$ and $\sigma_{\rm V}$ are estimated by comparing PV diagrams extracted from both the model and observed cubes (Fig.\,\ref{fig:Models}, bottom panels). The major-axis PV diagram is well reproduced by our axisymmetric kinematic model: the residuals are smaller than $\sim$2$\sigma$. The width of the \ci\ line profiles is largely driven by resolution effects but it can be used constrains the intrinsic gas velocity dispersion. Near the \ci\ center, the observed PV diagram shows asymmetric \ci\ profiles with a tail of emission towards the systemic velocity. These asymmetric profiles are well reproduced by our disc model, indicating that they are probably due to unresolved rotation. This is the typical effect of beam smearing in poorly resolved galaxy discs \citep[e.g.,][]{Lelli2010}. Channel maps are also individually inspected (Fig.\,\ref{fig:ChanMaps}): the \ci\ emission progresses along the major axis of the dust component, which is natural if both dust and \ci\ lie in a rotating disc. In particular, the estimated values of center, PA, and $i$ are consistent with the dust continuum, providing an independent constraint to the model.

Given the relatively low resolution and S/N ratio, $^{\rm 3D}$BAROLO cannot be run blindly leaving all the parameters free. We use $V_{\rm rot}$ and $\sigma_{\rm V}$ as free parameters, but keep the others fixed to the values estimated with GALMOD. The fit is performed within the Boolean mask described in Sect.\,\ref{sec:ALMA}. The best-fit values do not strongly depend on the mask definition, but they do depend on the initial estimates. The best-fit values of $V_{\rm rot}$ show a weak dependence and cluster around 310 \kms. The best-fit values of $\sigma_{\rm V}$ show a stronger dependence on the initial estimates, but they are always smaller than $\sim$30 \kms. Considering that the \ci\ cube has low S/N ratio and independent channels with a width of 50 \kms, the minimum detectable velocity dispersion is $\sim50/2.35 \simeq 21$, so we consider $\sigma_{\rm V}\lesssim30$ \kms as an hard upper limit. These values of $V_{\rm rot}$ and $\sigma_{\rm V}$ are in agreement with the previous visual estimates from GALMOD. To estimate errors on each variable, we re-run $^{\rm 3D}$BAROLO leaving such variable free and keeping all the others fixed to their best-fit values. The final parameters are given in Table\,\ref{tab:res}.

One may wonder whether the \ci\ velocity gradient may be explained by a gas outflow or a galaxy merger. The outflow scenario implies that (1) the gas is ejected along the major axis of the dust/stellar component rather than the minor axis (see Fig.\,\ref{fig:Atlas}), and (2) the PA of the outflow is tilted with respect to the synchrotron lobes, which are generally tracing the AGN ionization cone \citep[e.g.,][]{Drouart2012}. Both facts seem unlikely. 

The merger origin for the \ci\ velocity gradient cannot be definitively ruled out, but we disfavour this scenario given the high symmetry of the observed PV diagram. A kinematic merger model would have twice as many free parameters compared to a disc model and would need to be fine-tuned: (1) two merging \ci\ components should have roughly the same mass, size, and velocity dispersion, (2) they should be aligned along the major axis of the dust continuum and have the same projected distance from its centroid (see Fig.\,\ref{fig:ChanMaps}), and (3) they should be well separated in velocity by $\sim$500 km\,s$^{-1}$ and lie very close on the sky ($\lesssim$0.5$''$), but they should not overlap at any spatial position after beam convolution, otherwise we should observe double-peaked \ci\ profiles (contrary to what is seen in Fig.\,\ref{fig:Models}). Naturally, PKS\,0529-549 may still be a late-stage merger with an inner rotating disc, as observed in some starburst galaxies at $z\simeq0$ \citep[e.g.,][]{Kregel2001, Weaver2018}. Here we simply point out that the observed \ci\ velocity gradient is probably due to rotation, allowing the following dynamical analysis.

\subsection{Mass budget}

\citet{deBreuck2010} measured the rest-frame $H$-band luminosity ($L_{\rm H}$) and stellar mass ($M_{\star}$) of 70 radio galaxies. For PKS\,0529-549, they give $L_{\rm H} \simeq 4 \times 10^{11}$ L$_{\odot}$ and $M_{\star} \simeq 3 \times 10^{11}$ M$_{\odot}$. The stellar mass is derived by fitting the $Spitzer$ SED and assuming (i) a formation redshift of 10, (ii) an exponentially declining star formation history (SFH), and (iii) a \citet{Kroupa2001} initial mass function (IMF). These assumptions are sensible for radio galaxies in a statistical sense \citep{RoccaVolmerange2004}. For individual galaxies, however, the SFH remains uncertain. PKS\,0529-549 is indeed bursting at $z\simeq2.6$ since the estimated star formation rate (SFR) is $\sim$1000 M$_{\odot}$ yr$^{-1}$ (T.\,Falkendal et al. 2018, in prep.). For reference, galaxies with $M_{\star}\gtrsim10^{11}$ M$_\odot$ at $z\simeq2-3$ have SFRs of the order of $\sim$100 M$_{\odot}$ yr$^{-1}$ \citep[e.g.,][]{Bisigello2018}: PKS\,0529-549 lies well above the mean SFR$-M_{\star}$ relation (the so-called ``star-forming main sequence'').

The effect of starbursts in high-$z$ radio galaxies has been studied by \citet{Drouart2016} in eleven objects by modeling their full SED, from ultraviolet to NIR wavelengths. Unfortunately, their sample does not include PKS\,0529-549 but it does include 4C\,41.17, which is analogous to PKS\,0529-549 in several aspects. In particular, the rest-frame optical spectrum of 4C\,41.17 is strikingly similar to the one of PKS\,0529-549 (A. Man et al. 2018, in prep.). \citet{Drouart2016} finds that 4C\,41.17 is experiencing a recent starburst ($\sim$30 Myr) which makes up $\sim$92$\%$ of the bolometric luminosity and forms $\sim$44$\%$ of the total stellar mass. Using this analogy and considering the large uncertainties, we adopt a conservative errorbar on the stellar mass of PKS\,0529-549: $M_\star = (3.0\pm2.0) \times 10^{11}$ $M_{\odot}$. This errorbar does not include systematic uncertainty due to the IMF: a Salpeter IMF would increase the stellar mass by $\sim$60$\%$.

The gas mass is given by three main mass components: atomic hydrogen ($M_{\hi}$), molecules ($M_{\rm mol}$), and the warm ionized medium ($M_{\rm WIM}$). Unfortunately, \hi\ emission is currently not accessible at high $z$. However, it is likely that the central parts of massive starburst galaxies are dominated by H$_2$ rather than \hi, as observed in nearby interacting galaxies \citep{Kaneko2017}. Considering the observed \ci$(2-1)$ flux of 2.1 Jy\,\kms\ and the theoretical relations from \citet{Papadopoulos2004}, we estimate $M_{\rm mol}\simeq 4 \times 10^{10}$ M$_{\odot}$. This \ci$-$to$-$H$_2$ conversion assumes the [C/H$_2$] abundance of the local starburst galaxy M82 and is uncertain by about a factor of 2 (see also A. Man et al. 2018, in prep.). Finally, the mass of the WIM is clearly negligible: \citet{Nesvadba2017} estimate $2.4\times10^{9}$ M$_{\odot}$ for a fiducial electron density of 500 cm$^{-3}$. Hence, the total baryonic mass of PKS\,0529-549 is $M_{\rm bar} = (3.4\pm2.0) \times 10^{11}$ $M_{\odot}$.

\begin{table}
\caption{Results of the 3D dynamical analysis.}
\begin{center}
\begin{tabular}{l c}
\hline
Parameter                 & Value \\
\hline
$x_0$ (J2000)             & 05$^{\rm h}$ 30$^{\rm m}$ 25.44$^{\rm s}$ $\pm$ 0.01$^{\rm s}$ \\
$y_0$ (J2000)             & $-$54$^{\circ}$ 54$'$ 23.2$''$ $\pm$ 0.1$''$  \\
$z$ (from $V_{\rm sys}$)  & $2.570\pm0.001$ \\
PA ($^{\circ}$)           & $75\pm12$ \\
$i$ ($^{\circ}$)          & $50\pm4$ \\
$V_{\rm rot}$ (\kms)      & $310\pm50$ \\
$\sigma_{\rm V}$ (\kms) & $\lesssim$30\\
\hline
$M_{\star}$ (10$^{11}$ M$_{\odot}$)   & 3.0$\pm$2.0$^{a}$\\
$M_{\rm mol}$ (10$^{11}$ M$_{\odot}$) & 0.4$\pm$0.2$^{b}$\\
$M_{\rm WIM}$ (10$^{11}$ M$_{\odot}$) & 0.02$\pm$0.02$^{c}$\\
$M_{\rm bar}$ (10$^{11}$ M$_{\odot}$) & 3.4$\pm$2.0\\
$M_{\rm 4\,kpc}$ (10$^{11}$ M$_{\odot}$) & 0.9$\pm$0.3 \\
$M_{\rm 8\,kpc}$ (10$^{11}$ M$_{\odot}$) & 1.8$\pm$0.6 \\
\hline
\end{tabular}
\end{center}
References: $^{a}$\citet{deBreuck2010}, $^{b}$Man et al. (2018, in prep.), $^{c}$\citet{Nesvadba2017}.
\label{tab:res}
\end{table}

The dynamical mass within a radius $R$ is given by
\begin{equation}
M_{\rm R}  = \varepsilon \, \frac{ R\, V_{\rm rot}^{2} }{G}
\end{equation}
where $G$ is the gravitational constant and $\varepsilon$ is a parameter of the order of unity that depends on the 3D geometry of the \textit{total} mass distribution. For simplicity, we adopt $\varepsilon = 1$ (spherical geometry). For local disc galaxies, \citet{McGaugh2005} estimates $\varepsilon \simeq 0.8$ , so one would infer a slightly smaller dynamical mass. For PKS\,0529-549 we measure $V_{\rm rot} = 310 \pm 50$ km~s$^{-1}$ within the deconvolved radius of the \ci\ disk ($R\simeq0.45''\simeq4$ kpc), thus $M_{\rm 4\, kpc} = (0.9 \pm 0.3) \times10^{11}$ M$_{\odot}$.

The value of $M_{\rm 4\,kpc}$ is consistent with $M_{\rm bar}$ within the errors albeit it appears somewhat smaller. It is likely, however, that the stellar component is more extended than the \ci\ disc. Unfortunately, the available NIR images have poor spatial resolution, but they tentatively indicate that the stellar component may be twice as extended as the \ci\ disc \citep[cf. with Figure 5 of][]{Broderick2007}. If we extrapolate the measured rotation velocity out to 8 kpc, we get $M_{\rm 8\,kpc} = (1.8\pm0.6) \times 10^{11}$ M$_{\odot}$ which is closer to $M_{\rm bar}$. Alternatively, one may question the estimated inclination. If the \ci\ disc were oriented more face-on ($i\simeq30^{\circ}$), the rotation velocity would increase to $\sim475$ \kms\ and give $M_{\rm 4\,kpc}\simeq 2 \times 10^{11}$ M$_{\odot}$. These values of $i$ and $V_{\rm rot}$, however, are difficult to recoincile with the observed \ci\ map (Fig.\,\ref{fig:Models}, top panels).

In conclusion, the baryonic and dynamical masses of PKS\,0529-549 are consistent within the uncertainties. This suggests that PKS\,0529-549 has little (if any) dark matter in its central parts. This is comparable to other massive galaxies at similar redshifts \citep{Toft2012, Wuyts2016} as well as massive late-type galaxies \citep{Noordermeer2007, Lelli2016b} and early-type galaxies \citep{Cappellari2016} at $z\simeq0$. The latter ones are commonly thought to be the descendants of high-$z$ radio galaxies like PKS\,0529-549 \citep[e.g.,][]{Pentericci2001}.

\section{Discussion}\label{sec:Disc}

\subsection{Comparison with other high-$z$ galaxies}\label{sec:compa}

The first IFS surveys at $z\simeq1-3$ suggested that 1/3 of star-forming galaxies were rotation-dominated systems, whereas the remaining 2/3 were either merging systems or dispersion-dominated objects \citep{Forster2009, Gnerucci2011}. Subsequently, it was acknowledged that beam-smearing effects can play a major role in assessing the intrinsic galaxy dynamics: the most recent IFS surveys now indicate that at least $\sim$80$\%$ of massive star-forming galaxies ($10^{9} < M_{\star}/M_{\odot} < 10^{11}$) host rotating gas discs at $z\simeq1-3$ \citep{Wisnioski2015, Stott2016}. 

While rotating gas discs are now thought to be ubiquitous at $z\simeq1-3$, their detailed properties are still debated. The general picture is that high-$z$ discs are more turbulent than their local analogues \citep{Forster2009, Lehnert2009, Gnerucci2011}: $\sigma_{\rm V}$ is thought to increase systematically with $z$, while $V_{\rm rot}/\sigma_{\rm V}$ decreases \citep{Wisnioski2015, Stott2016}. \citet{DiTeodoro2016}, however, point out that high-$z$ discs are poorly resolved and beam-smearing effects can lead to systematic over-estimates of $\sigma_{\rm V}$ and under-estimates of $V_{\rm rot}$, unless they are properly modelled in 3D. For PKS\,0529-549 we find $V_{\rm rot}/\sigma_{\rm V} \gtrsim 10$, which is comparable to CO and \hi\ discs in spiral galaxies at $z=0$ \citep{Leroy2008, Mogotsi2016}. Thus, the \ci\ disc of PKS\,0529-549 is not particularly turbulent. This is remarkable considering that PKS\,0529-549 is forming stars at $\sim$1000 M$_{\odot}$ yr$^{-1}$ and hosts a powerful radio-loud AGN, so a large amount of energy should be injected into the inter-stellar medium \citep[e.g.,][]{Lehnert2009}.

The gas phase may be important in assessing the degree of turbulence of a galaxy disc. In general, ionized gas has higher $\sigma_{\rm V}$ than neutral gas due to thermal broadening, but is also more strongly affected by stellar and AGN feedback. Thus, it is possible that some high values of $\sigma_{\rm V}$ in ``turbulent'' discs are actually due to undetected non-circular motions and/or outflows in the ionized gas component. For example, the \oiii\ velocity field in Fig.\,\ref{fig:Atlas} could be interpreted as a turbulent disc if radio images and/or \ci\ kinematics were not available. It is possible, indeed, to fit the \oiii\ cube with a rotating disc model, following the same procedures described in Sect.\,\ref{sec:disc} for the \ci\ cube. This gives $V_{\rm rot}\simeq100$ \kms and $\sigma_{\rm V}\simeq 300$ \kms, leading to $V_{\rm rot}/\sigma_{\rm V}\simeq0.3$. Large IFS surveys find that ``main-sequence'' galaxies at $z\simeq2-3$ have $V_{\rm rot}/\sigma_{\rm V} \simeq 1-4$ when using H$\alpha$ or \oiii\ lines \citep{Wisnioski2015}. The lower value of PKS\,0529-549 corroborates the outflow interpreation for the \oiii\ line. However, it also raises the question of how much gas outflows may affect the mean value of $V_{\rm rot}/\sigma_{\rm V}$ in large IFS surveys. Clearly, a multiwavelength approach is needed to have a complete picture of the gas dynamics of high-$z$ galaxies.

\subsection{The baryonic Tully-Fisher relation}\label{sec:BTFR}

A key constraint for galaxy formation models is represented by the emergence and evolution of the Tully-Fisher (TF) relation \citep{Tully1977}. Over the past years, both the stellar-mass TF relation (using $M_\star$) and the baryonic TF relation (using $M_{\rm bar} = M_\star + M_{\rm gas}$) have been investigated up to $z\simeq3$ with contradicting results. Some studies find significant evolution \citep{Tiley2016, Ubler2017}, while others do not \citep{Miller2011, Miller2012, DiTeodoro2016, Pelliccia2017}. According to \citet{Turner2017}, these discrepancies are due to selection effects since different authors use different criteria to define ``disc-dominated'' galaxies. Data quality must also play a major role. Different authors use different techniques to derive rotation velocities, velocity dispersions, and stellar masses. The latter ones can be particularly uncertain when comparing objects at different redshifts due to systematic effects. For example, \citet{Turner2017} find that the normalization of the stellar-mass TF relation varies by $\sim$0.1 dex from $z=0$ to $z\simeq3$. This offset is comparable to the uncertainty in the absolute calibration of stellar masses at $z=0$ from $Spitzer$ [3.6] photometry \citep[e.g.,][]{Lelli2016b}, so it is hard to tell whether the stellar-mass TF relation actually evolves. The situation is exacerbated when considering the baryonic TF relation: the gas fraction of galaxies ($M_{\rm gas}/M_{\star}$) likely evolves with redshift, but direct estimates of gas masses are rarely available. 

\begin{figure}
\includegraphics[width=0.45\textwidth]{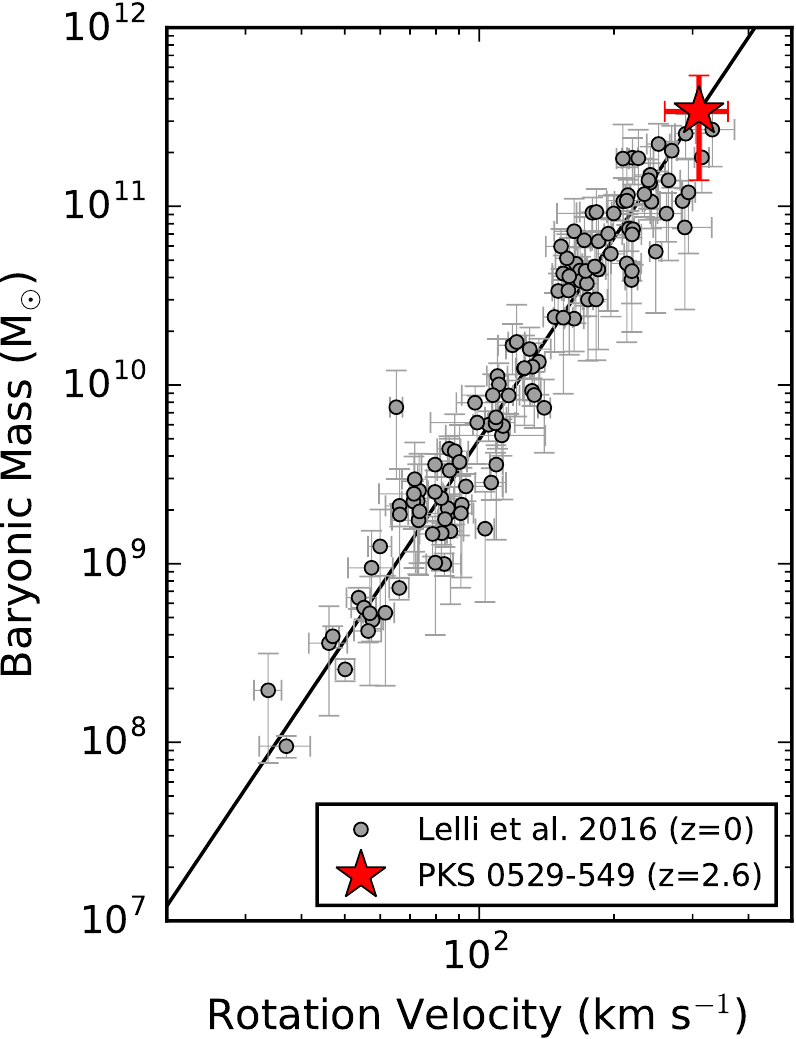}
\caption{The location of PKS\,0529-549 on the local BTFR from the SPARC sample \citep{Lelli2016b, Lelli2016a}.}
\label{fig:BTFR}
\end{figure}
In this context, PKS\,0529-259 is particularly interesting because we have estimates of both $M_{\star}$ and $M_{\rm mol}$, other than $V_{\rm rot}$ and $\sigma_{\rm V}$. Figure\,\ref{fig:BTFR} shows that PKS0529-259 lies on the local baryonic TF relation from the SPARC sample \citep{Lelli2016b, Lelli2016a}. This baryonic TF relation is consistent with previous calibrations \citep[e.g.,][]{McGaugh2005}, but has the advantage of using homogeneous photometry at $Spitzer$ [3.6], which greatly reduces the uncertainties in $M_\star$. The stellar masses are calculated assuming $M_\star/L_{[3.6]}= 0.5$ $M_{\sun}/L_{\sun}$ for all galaxies, as expected from stellar population synthesis models with a \citet{Kroupa2001} IMF (same as for PKS\,0529-549). The rotation velocities are measured along the flat part of the rotation curve ($V_{\rm flat}$), which is probed by deep \hi\ observations \citep[see][for details]{Lelli2016b, Lelli2016a}. In the case of PKS\,0529-549, $V_{\rm rot}$ is an intensity-weighted estimate over the semi-major axis since the \ci\ emission is resolved with $\sim$2 beams. Thus, one may wonder whether we are probing $V_{\rm flat}$. Local galaxies with similar masses as PKS\,0529-549 typically have rotation curves that peak at very small radii ($R\lesssim1$ kpc) and decline by $\sim20-30\%$ before reaching $V_{\rm flat}$ \citep{Noordermeer2007}. If the rotation curve of PKS\,0529-549 has a similar shape, then our intensity-weighted value should not differ from $V_{\rm flat}$ by more than $\sim20-30\%$. In general, the adherence of PKS\,0529-549 to the baryonic Tully-Fisher relation seems solid against systematic effects. Clearly, a single object cannot be used to infer general conclusions. However, Figure\,\ref{fig:BTFR} suggests that some massive galaxies can be in place and kinematically relaxed at $z\simeq2.6$, when the Universe was only $\sim$2.5 Gyr old.

\section{Conclusions}

We presented ALMA data of the $\ci\,(2-1)$ line for the starburst galaxy PKS\,0529-549 at $z\simeq2.6$. With only 5 minutes of on-source integration time, ALMA spatially resolves the \ci\ distribution and kinematics. We also analysed SINFONI data \citep{Nesvadba2017}, probing the ionized gas kinematics, and ATCA radio continuum images \citep{Broderick2007}, probing synchrotron emission from a central AGN. Our results can be summarized as follows:
\begin{itemize}
 \item Both \ci\ and \oiii\ display regular velocity gradients, but their systemic velocities and kinematic PAs differ by $\sim$300 \kms\ and $\sim$30$^{\circ}$, respectively. The \ci\ is consistent with a rotating disc, being aligned with both the stellar and dust components, while the \oiii\ likely traces an outflow, being aligned with two AGN-driven radio lobes.
 \item The \ci\ cube is well reproduced by a 3D disc model with $V_{\rm rot} \simeq 310$ \kms\ and $\sigma_{\rm V}\lesssim 30$ \kms. This gives $V_{\rm rot}/\sigma_{\rm V} \gtrsim 10$ similar to local spiral galaxies, indicating that the \ci\ disc of PKS\,0529-549 is not particularly turbulent.
 \item The dynamical mass within 8 kpc is $\sim1.8\times10^{11}$ M$_{\odot}$. This is comparable to the baryonic mass within the errors, implying that baryons dominate over dark matter in the central parts. This is similar to massive galaxies at $z\simeq0$.
 \item PKS\,0529-549 lies on the local baryonic Tully-Fisher relation, suggesting that some massive galaxies can be already in place and kinematically relaxed at $z\simeq2.6$.
\end{itemize}
This study demonstrates the potential of the \ci\ line to trace galaxy dynamics at high-$z$. It also highlights the importance of multiwavelength observations to properly interpret gas kinematics during the early stages of galaxy formation.

\section*{Acknowledgements}

This paper makes use of the following ALMA data: ADS/JAO.ALMA$\#$2013.1.00521.S. ALMA is a partnership of ESO (representing its member states), NSF (USA) and NINS (Japan), together with NRC (Canada), MOST and ASIAA (Taiwan), and KASI (Republic of Korea), in cooperation with the Republic of Chile. The Joint ALMA Observatory is operated by ESO, AUI/NRAO and NAOJ.




\bibliographystyle{mnras}
\bibliography{PKS0529}

\bsp	
\label{lastpage}
\end{document}